\documentclass[12pt,a4paper,oneside,onecolumn]{article}
\usepackage[dvips]{epsfig}
\usepackage{fancyheadings}
\usepackage{fancybox}
\usepackage{graphicx}
\usepackage{subfigure}
\usepackage{amsmath}

\graphicspath{
  {./fig/}
    }

\def\bm{{\bf m}}

\def\ell{{h}}

\title{Coarse-Graining and Renormalization of Atomistic
Binding Relations and Universal Macroscopic Cohesive Behavior}

\date{}
\author{
  O.~Nguyen and M.~Ortiz\\
  Graduate Aeronautical Laboratories\\
  California Institute of Technology\\
  Pasadena CA 91125, USA
  }
\begin{document}

\maketitle

\abstract{We present two approaches for coarse-graining
interplanar potentials and determining the corresponding
macroscopic cohesive laws based on energy relaxation and the
renormalization group. We analyze the cohesive behavior of a
large---but finite---number of interatomic planes and find that
the macroscopic cohesive law adopts a universal asymptotic form.
The universal form of the macroscopic cohesive law is an
attractive fixed point of a suitably-defined
renormalization-group transformation.}

\section{Introduction}

Cohesive theories of fracture are predicated on a direct
description of the physical processes which lead to separation
and the eventual formation of a free surface. The development of
cohesive theories rests on a detailed physical understanding of
the operative fracture mechanisms, which are often complex and
cut across multiple lengthscales, especially where ductile
fracture is concerned. Cleavage fracture, by way of contrast,
entails the simple separation of atomic planes and is, therefore,
governed by interplanar potentials which are amenable to an
effective first-principles atomistic characterization. For
instance, Jarvis {\it et al.} \cite{JarvisHayesCarter2001} have
recently calculated the cohesive behavior of (111) planes in fcc
aluminum, and of Al${}_2$O${}_3$ cleavage planes, using GGA
density functional theory; and Park and Kaxiras
\cite{ParkKaxiras2001} have carried out ab-initio simulations of
hydrogen embrittlement in aluminum and calculated generalized
stacking-fault energies as a function of interplanar separation
and sliding.

First-principles interplanar potentials are characterized by peak
stresses of the order of the theoretical strength of the crystal.
In addition, the crystal loses its bearing capacity after an
interplanar separation of only a few angstroms. Moreover, the
integration of first-principles interplanar potentials into
engineering calculations necessitates full atomistic resolution
in the vicinity of the crack tip, which is often unfeasible or
impractical. This disconnect between atomistic and engineering
descriptions begs a number of fundamental question, to wit: What
is the proper way to \emph{coarse-grain} a cohesive description?,
and: What is the macroscopic form of the cohesive law after
coarse-graining?

In this paper we address these issues by investigating the
cooperative behavior of a large number of interatomic planes
forming a \emph{cohesive layer}. We employ two main approaches in
this investigation: relaxation and the renormalization group.
Relaxation or \emph{weak convergence} methods are concerned with
the determination of the macroscopic behavior of materials
characterized by a non-convex energy function. These materials
often develop fine microstructure in response to imposed
deformations. Truskinovsky {\it et al.}
\cite{DelpieroTruskinovsky1998, PuglisiTruskinovsky2000,
DelpieroTruskinovsky2001}, and Braides {\it et al.}
\cite{BraidesDalMasoGarroni1999}, have pioneered the application
of these methods to fracture. However, the full relaxation of a
cohesive potential yields the trivial result that the effective
cohesive potential is identically equal to zero. The chief
difference between the analysis pursued here and full relaxation
is that, at zero temperature, we seek energy minimizers of
large---\emph{but finite}---collections of interatomic planes. In
this limit we find that, for a broad class of interplanar
potentials, the macroscopic cohesive law adopts a \emph{universal
form} asymptotically.

We show that this universality of the macroscopic cohesive
behavior is amenable to a renormalization-group interpretation.
The normalization group which coarse-grains the cohesive behavior
is somewhat nonstandard and has to be crafted carefully, e.~g.,
so as to preserve the surface energy and the elasticity of the
lattice. The universal form of the macroscopic cohesive law is
precisely an attractive \emph{fixed point} of the
renormalization-group transformation.

\section{Problem formulation}

\begin{figure}[h]
  \begin{center}
    \epsfig{file=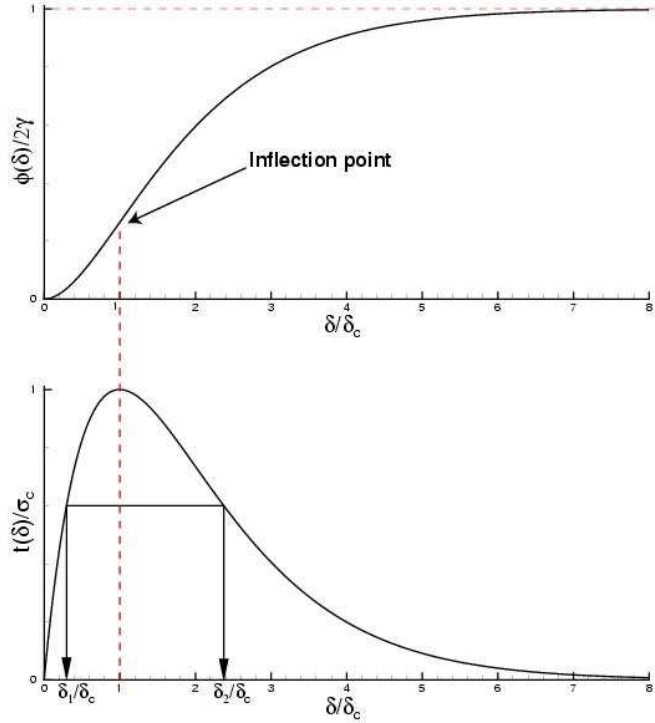,height=4.in}
    \caption[]{Interplanar potential and corresponding cohesive or
      binding law.}
    \label{fig:Potential}
  \end{center}
\end{figure}

We consider a macroscopic cohesive crack opening symmetrically
(mode I) and undergoing quasistatic growth. We denote by $d$ the
interplanar distance, $\delta$ the opening displacement across an
interatomic plane, and $t$ the corresponding cohesive traction.
These latter variables are presumed related by a known cohesive or
binding law $t(\delta)$, which derives from an interplanar
potential $\phi(\delta)$ through the relation
\begin{equation}
t(\delta) = \phi'(\delta)
\end{equation}
Here and subsequently, a prime denotes differentiation of a
function of a single variable. For simplicity, we shall assume
throughout that the atomistic binding law $t(\delta)$ rises
monotonically from zero at $\delta=0$ to a peak value $\sigma_c$
at $\delta=\delta_c$, and subsequently decreases monotonically to
zero, Fig.~\ref{fig:Potential}. Correspondingly, the cohesive potential
$\phi(\delta)$ is convex in the interval $0 \leq \delta <
\delta_c$, has an inflection point at $\delta=\delta_c$, is
concave for $\delta > \delta_c$ and asymptotes to twice the
surface energy, $2\gamma$, as $\delta\to\infty$. In addition, we
shall assume that $\phi(\delta)$ is smooth and analytic at
$\delta=0$, with Taylor expansion:
\begin{equation}\label{Taylor}
\phi \sim \frac{C}{2} \delta^2 + o(\delta^2)
\end{equation}
for some constant $C$.
The value of $C$ can be readily deduced
from the elastic moduli $c_{ijkl}$ of the crystal. To this end,
let $\bm$ be the unit normal to the plane of the crack, and apply
a small and uniform opening displacement to all interatomic
planes. Evidently, the energy per unit volume of the crystal
follows from the cohesive potential as $(C/2d) \delta^2$
asymptotically as $\delta \to 0$. On the other hand, the strain
tensor of the crystal is $\epsilon_{ij} = (\delta/d) m_im_j$, and
the corresponding energy is $(\delta^2/2d^2) c_{ijkl} m_i m_j m_k
m_l$. Equating both energies yields the identity:
\begin{equation}
C = \frac{1}{d} c_{ijkl} m_i m_j m_k m_l
\end{equation}

Next imagine that the atomistic description is coarse-grained,
e.~g., by the quasi-continuum method
\cite{TadmorOrtizPhillips1996}, or by a passage to the continuum
limit, or by some other suitable means. Let $\bar{d}$ denote the
spatial resolution of the coarse-grained description. For
instance, in quasicontinuum or in engineering finite-element
simulations $\bar{d}$ measures the local element size. The
corresponding effective cohesive law may be obtained by analyzing
the behavior of a \emph{cohesive layer} of thickness $\bar{d}$
and containing $N = \bar{d}/d$ atomic planes. The cohesive layer
is taken through a total opening displacement $\bar{\delta}$
resulting in a macroscopic traction $\bar{t}$. The chief objective
of the analyses that follow is to determine the macroscopic
cohesive law $\bar{t}(\bar{\delta})$ in the limit of $N$ large
but finite. Equivalently, we may seek to determine the asymptotic
form of the macroscopic cohesive potential
$\bar{\phi}(\bar{\delta})$ such that
\begin{equation}
\bar{t}(\bar{\delta}) = \bar{\phi}'(\bar{\delta})
\end{equation}
in the same limit.

\section{Universal asymptotic form of the macroscopic cohesive
law at zero temperature} \label{ZeroTemperature}

At zero temperature, the crystal deforms so as to minimize its
total energy. The governing principle is, therefore, energy
minimization. Let $\delta_i \geq 0$, $i = 1,\dots, N$ be the
opening displacements of the interatomic planes in the cohesive
layer. Then, the total energy of the cohesive layer is:
\begin{equation}\label{Etot}
E^{\rm tot} = \sum_{i=1}^N  \phi(\delta_i)
\end{equation}
Let now $\bar{\delta}$ be the macroscopic opening displacement.
Then, the effective or macroscopic energy of the cohesive layer
follows from the constrained minimization problem:
\begin{eqnarray}
\bar{\phi}(\bar{\delta}) &=& \inf_{\{\delta_1, \dots, \delta_N\}}
\sum_{i=1}^N  \phi(\delta_i) \label{Energy} \\
\bar{\delta} &=& \sum_{i=1}^N \delta_i \label{Kinematics}
\end{eqnarray}
In conjunction with the kinematic constraints (\ref{Kinematics})
the stationarity of $E^{\rm tot}$ demands:
\begin{equation}\label{Equilibrium}
t(\delta_i) = \bar{t}(\bar{\delta}) = \text{constant}, \quad i =
1,\dots, N
\end{equation}
Thus, at equilibrium, all interplanar tractions must be equal to
the macroscopic traction.

We shall classify the possible states of an interatomic plane into
two categories or \emph{variants}, according as to whether the
opening displacement $\delta$ is in the range $0 \leq \delta <
\delta_c$, or in the range $\delta > \delta_c$. We shall
designate variants of the first kind as \emph{coherent}, and
variants of the second kind as \emph{decohered}. We may further
classify the states of a cohesive layer by the number $N_1$ of
coherent planes, or, equivalently, the number $N_2$ of decohered
planes, it contains. Since the function $t(\delta)$ is one-to-one
over the interval $[0,\delta_c)$, eq.~(\ref{Equilibrium}) demands
that the opening displacements of all coherent planes be equal at
equilibrium. Likewise, the opening displacements of all decohered
planes must be identical at equilibrium. Under these conditions
the macroscopic cohesive energy follows from the minimization
problem:
\begin{eqnarray}
\bar{\phi}(\bar{\delta}) &=& \inf_{\{(\delta_1, \delta_2), (N_1,
N_2) \}} \{ N_1
\phi(\delta_1) + N_2 \phi(\delta_2) \} \label{Energy2} \\
\bar{\delta} &=& N_1 \delta_1 + N_2 \delta_2 \label{Kinematics2}
\\
N &=& N_1 + N_2 \label{N} \\
0 &\leq& \delta_1 < \delta_c \\
\delta_2 &>& \delta_c
\end{eqnarray}
where $\delta_1$ and $\delta_2$ are the opening displacements in
the coherent and decohered planes, respectively. In addition, the
equilibrium equations (\ref{Equilibrium}) reduce to:
\begin{equation}\label{Equilibrium2}
t(\delta_1) = t(\delta_2) = \bar{t}(\bar{\delta})
\end{equation}
These relations are depicted geometrically in Fig.~\ref{fig:Potential}.

Next we proceed to determine the minimum energy states of a
cohesive layer by analyzing the cases $N_2 = 0, 1, 2, \dots$ in
turn. We begin by considering the case in which all planes are
coherent, corresponding to $N_1=N$ and $N_2=0$. Then, the
kinematic constraint (\ref{Kinematics}) gives $\delta_1 =
\bar{\delta}/N \equiv \delta$. Evidently, in the limit of $N \to
\infty$ $\delta$ tends to zero and, in view of
eqs.~(\ref{Taylor}) and (\ref{Energy2}), we obtain:
\begin{equation}\label{Energy3}
\bar{\phi}(\bar{\delta})_{\big| N_2=0} \sim \frac{\bar{C}}{2}
\bar{\delta}^2, \qquad \text{as } N \to \infty
\end{equation}
where
\begin{equation}
\bar{C} = \frac{C}{N}
\end{equation}
is an effective cohesive-layer stiffness.

Consider next the case of one decohered plane, $N_1=N-1$ and
$N_2=1$, whence (\ref{Kinematics2}) becomes:
\begin{equation}\label{Kinematics3}
(N-1) \delta_1 + \delta_2 = \bar{\delta}
\end{equation}
Solving for $\delta_1$ gives $\delta_1 = (\bar{\delta} -
\delta_2)/(N-1)$. In addition, since $\delta_i$ and $\bar{\delta}$
are required to be nonnegative, it follows that $\delta_1 \leq
\bar{\delta}/(N-1)$, and thus $\delta_1 \to 0$ as $N \to \infty$.
From this limit it additionally follows that $\delta_2 \to
\bar{\delta}$ in the same limit. Suppose now that $\bar{\delta}
\gg \delta_c$ and, hence, $\delta_2 \gg \delta_c$. Under these
conditions, $\phi(\delta_2) \sim 2\gamma$ and (\ref{Energy2})
reduces to:
\begin{equation}\label{Energy4}
\bar{\phi}(\bar{\delta})_{\big| N_2=1} \sim 2 \gamma, \qquad
\text{as } N \to \infty
\end{equation}
Since interactions beyond nearest neighbors are not taken into
account, an altogether identical analysis gives
\begin{equation}\label{Energy5}
\bar{\phi}(\bar{\delta})_{\big| N_2=k} \sim 2 k \gamma, \qquad
\text{as } N \to \infty
\end{equation}
for the case of $k$ decohered planes.

The macroscopic cohesive energy may now be expressed as
\begin{equation}\label{Energy6}
\bar{\phi}(\bar{\delta}) = \min_{0\leq k \leq N}
\bar{\phi}(\bar{\delta})_{\big| N_2=k}
\end{equation}
However, it follows from (\ref{Energy5}) that, asymptotically as
$N \to \infty$, multiple decohered planes are not energetically
possible at zero temperature, and only the cases $k=0$ and $k=1$
need be considered in (\ref{Energy6}). Therefore, the effective
cohesive potential is the lower envelop of the energies
(\ref{Energy3}) and (\ref{Energy4}), namely,
\begin{equation}\label{AsymptoticPhi}
\bar{\phi}(\bar{\delta}) = \min \{\frac{\bar{C}}{2}
\bar{\delta}^2, 2\gamma \} = \left\{
\begin{array}{cc}
(\bar{C}/2) \bar{\delta}^2, &
\text{if } \bar{\delta} < \bar{\delta}_c \\
2 \gamma, & \text{otherwise}
\end{array}
\right.
\end{equation}
where
\begin{equation}\label{DeltaCritical}
\bar{\delta}_c = 2 \sqrt{\frac{\gamma}{\bar{C}}} = 2
\sqrt{\frac{\gamma N}{C}}
\end{equation}
is a macroscopic critical opening displacement for the nucleation
of a single decohered plane. The corresponding macroscopic
cohesive law is
\begin{equation}
\bar{t}(\bar{\delta}) = \left\{
\begin{array}{cc}
\bar{C}\bar{\delta}, &
\text{if } \bar{\delta} < \bar{\delta}_c \\
0, & \text{otherwise}
\end{array}
\right.
\end{equation}
It is interesting to note that the peak macroscopic traction is:
\begin{equation}
\bar{\sigma}_c = \bar{C} \bar{\delta}_c = 2 \sqrt{\bar{C} \gamma}
= 2 \sqrt{\frac{C \gamma}{N}}
\end{equation}
We conclude the analysis by verifying that, in the decohered
regime, $\bar{\delta} > \bar{\delta}_c \sim \sqrt{N}$, and hence
$\bar{\delta} \gg \delta_c$ for sufficiently large $N$, as
supposed.

\begin{figure}[th]
  \begin{center}
    \epsfig{file=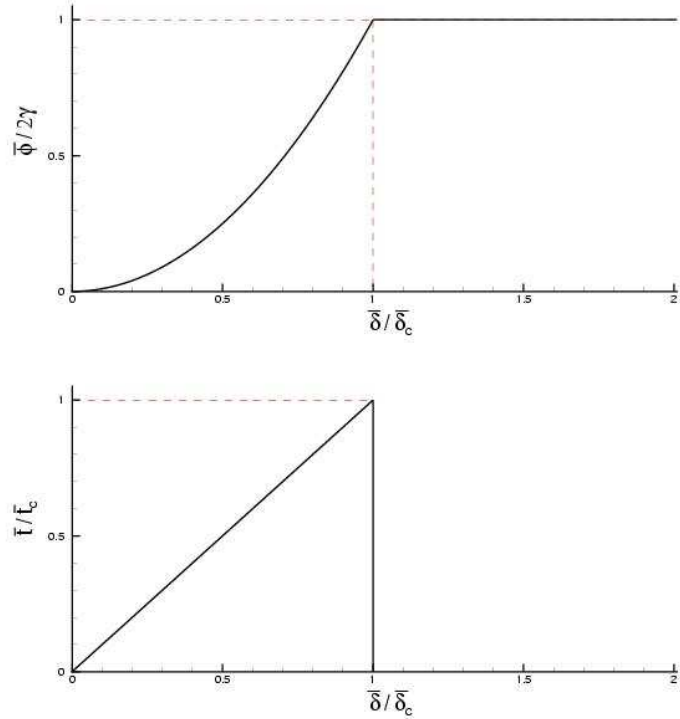,height=4.in}
    \caption[]{Universal asymptotic form of the macroscopic cohesive
      law at zero temperature.}
    \label{fig:PotentialRN}
  \end{center}
\end{figure}

These functions are shown in Fig.~\ref{fig:PotentialRN}. The macroscopic
cohesive potential is initially quadratic and subsequently
constant following the attainment of the critical macroscopic
opening displacement. Remarkably, the macroscopic critical
opening displacement and peak traction scale as: $\bar{\delta}_c
\sim \sqrt{N}$ and $\bar{\sigma}_c \sim 1/\sqrt{N}$,
respectively. Thus, for large $N$, it follows that the macroscopic
cohesive law entails much lower tractions, occurring at much
larger opening displacements, than the atomistic binding law. In
effect, the passage from the atomistic to the macroscopic scales
is accompanied by an expansion of the opening displacement axis
and a simultaneous compression of the traction axis. By
constrast, the macroscopic fracture energy, or critical
energy-release rate, $\bar{\phi}(\infty)$ remains invariant under
the transformation and is equal to the atomistic value $\phi(\infty)
= 2\gamma$.
It is also remarkable that, for the class of binding laws under
consideration, the asymptotic form (\ref{AsymptoticPhi}) of the
macroscopic cohesive law is \emph{universal}, i.~e., independent
of the atomistic binding law. Evidently, the parameters which
define the macroscopic cohesive law quantitatively, e.~g., the
surface energy $\gamma$ and the modulus $C$, are material
specific.

\begin{figure}[h]
  \begin{center}
    \epsfig{file=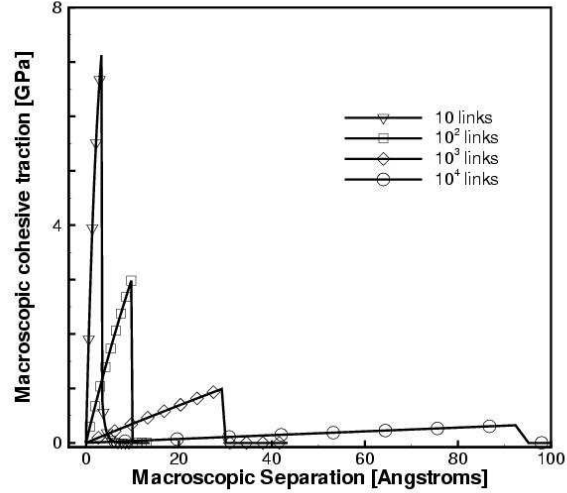,height=3.in}
    \caption[]{Numerically computed macroscopic traction {\it vs}
    opening displacement relation for an increasing number of
    atomic planes in the cohesive layer.}
    \label{fig:UBERtraction}
  \end{center}
\end{figure}

\begin{figure}[h!]
  \centering
  \subfigure[$N=10$]{
    \label{tenlinks}
    \includegraphics[width=.33\textwidth]{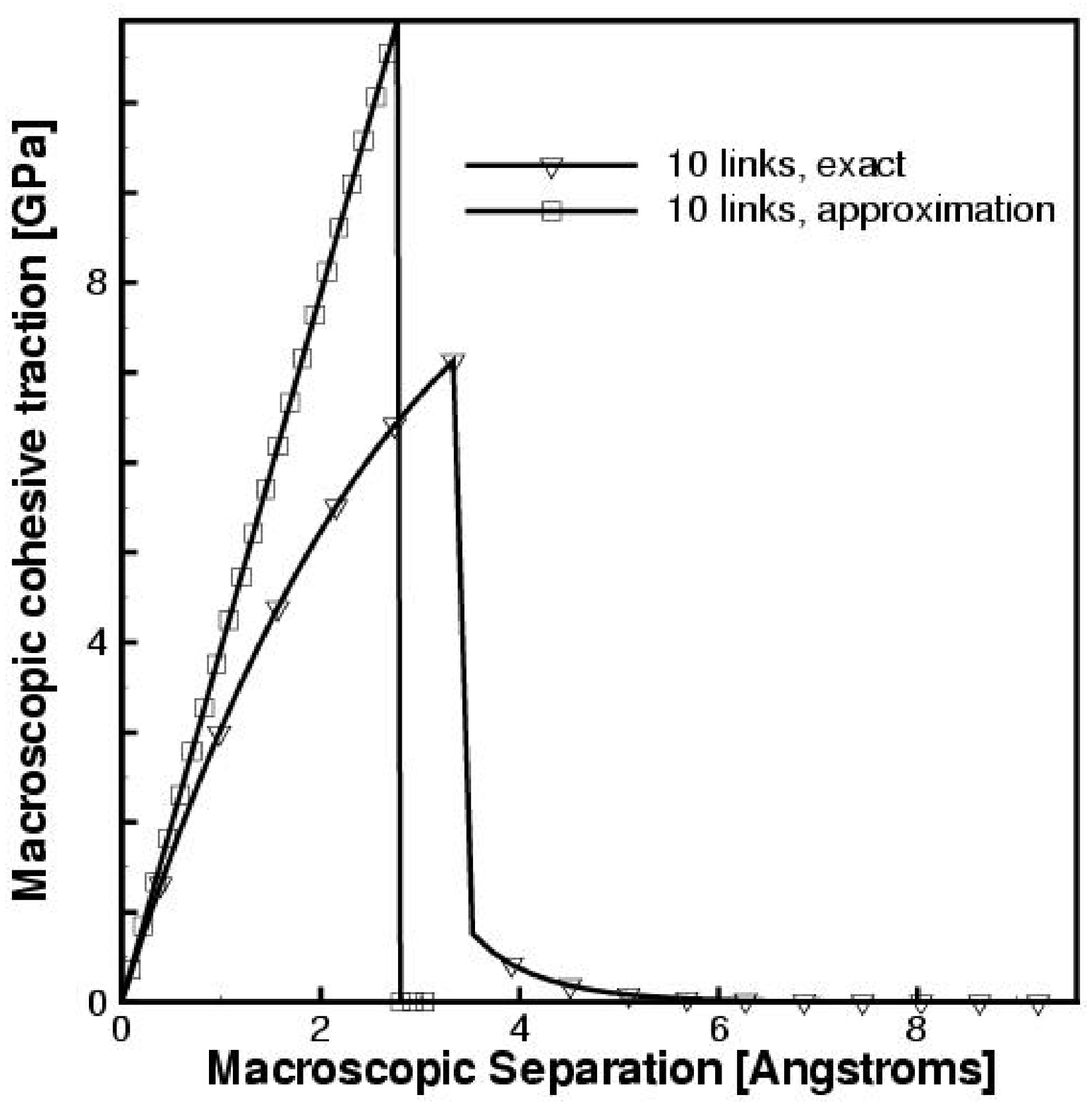}
    }
  \subfigure[$N=100$]{
    \label{hlinks}
    \includegraphics[width=.33\textwidth]{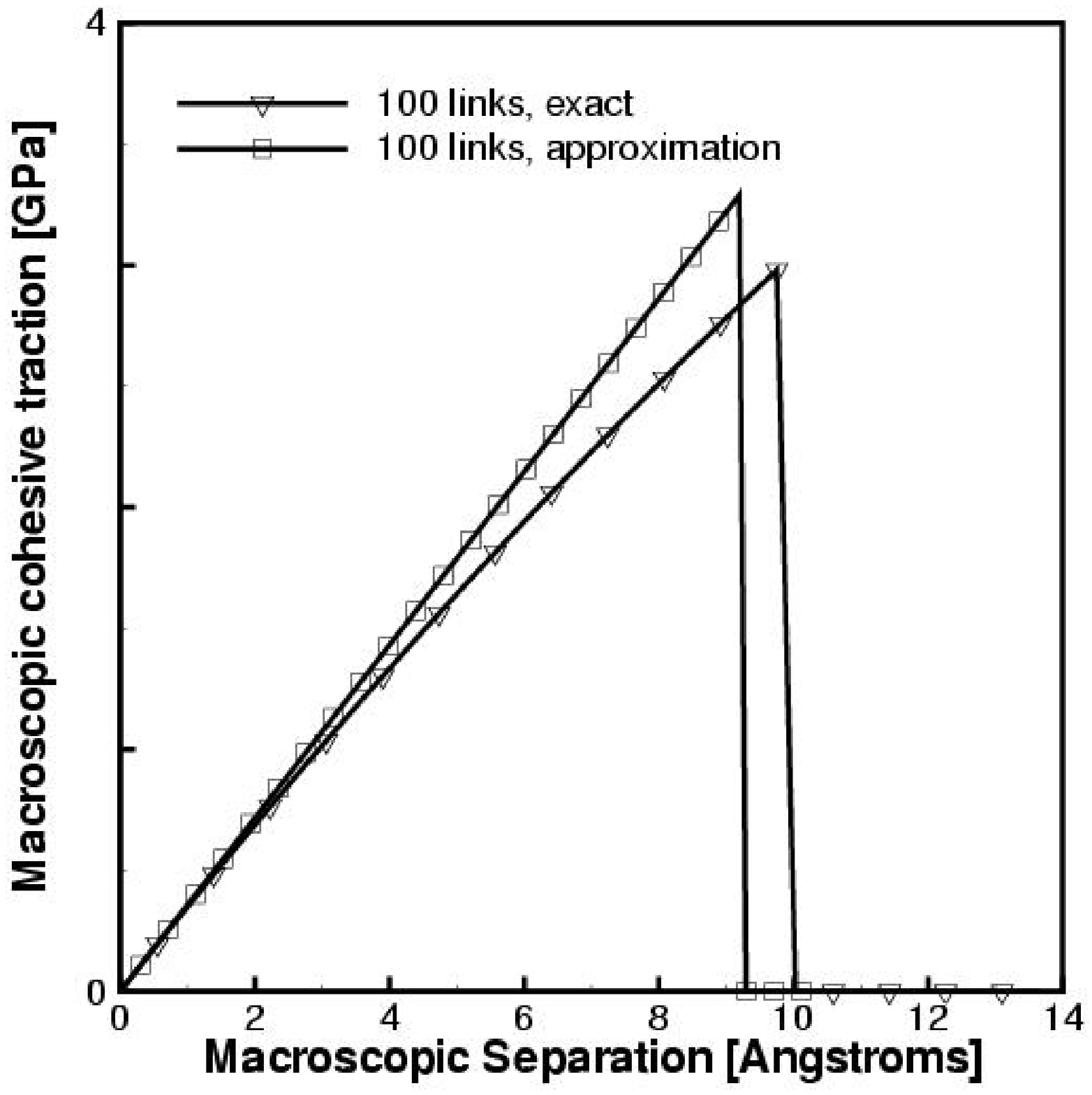}
    }
  \subfigure[$N=1000$]{
    \label{otlinks}
    \includegraphics[width=.33\textwidth]{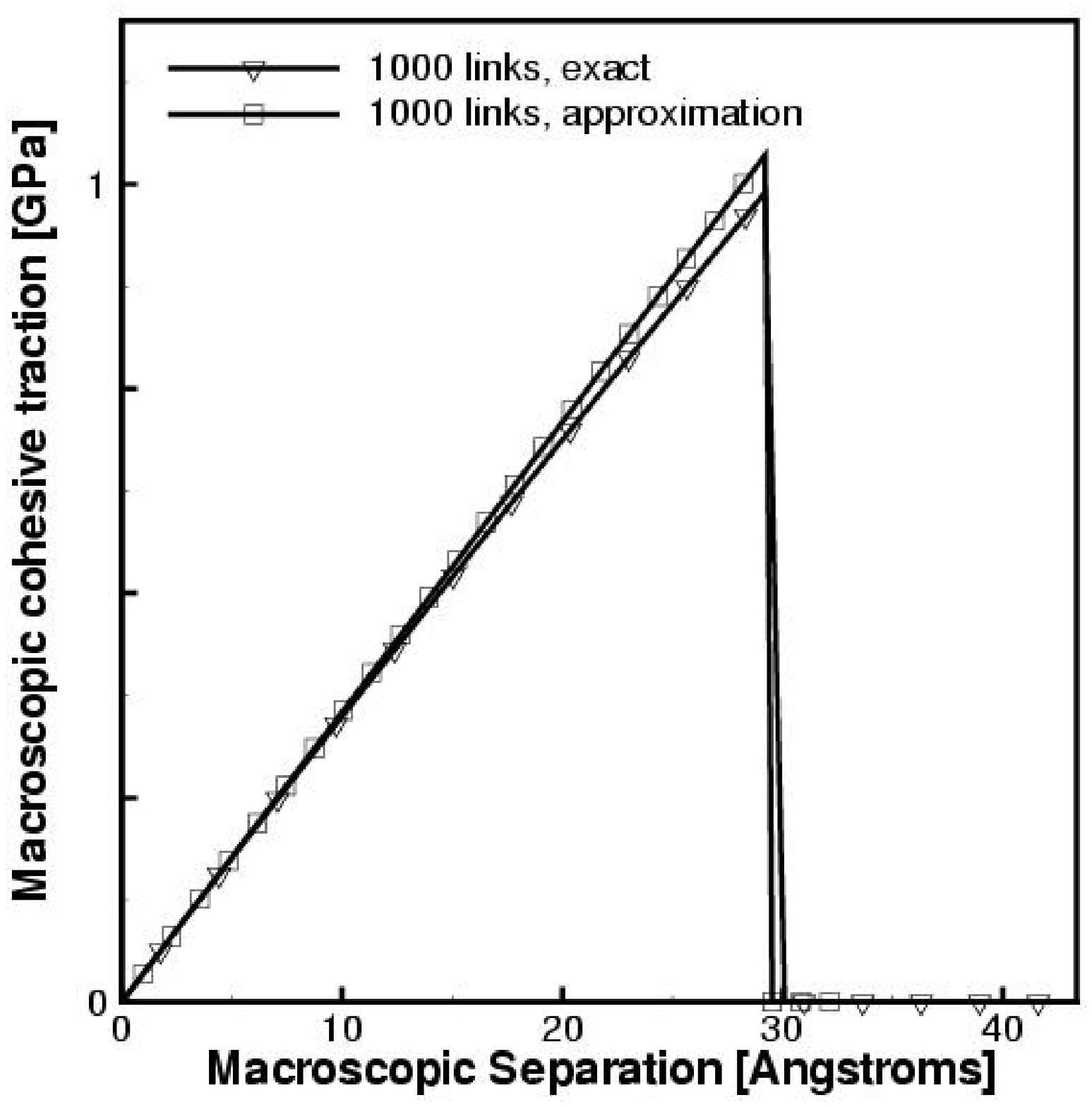}
    }
  \subfigure[$N=10^4$]{
    \label{ttlinks}
    \includegraphics[width=.33\textwidth]{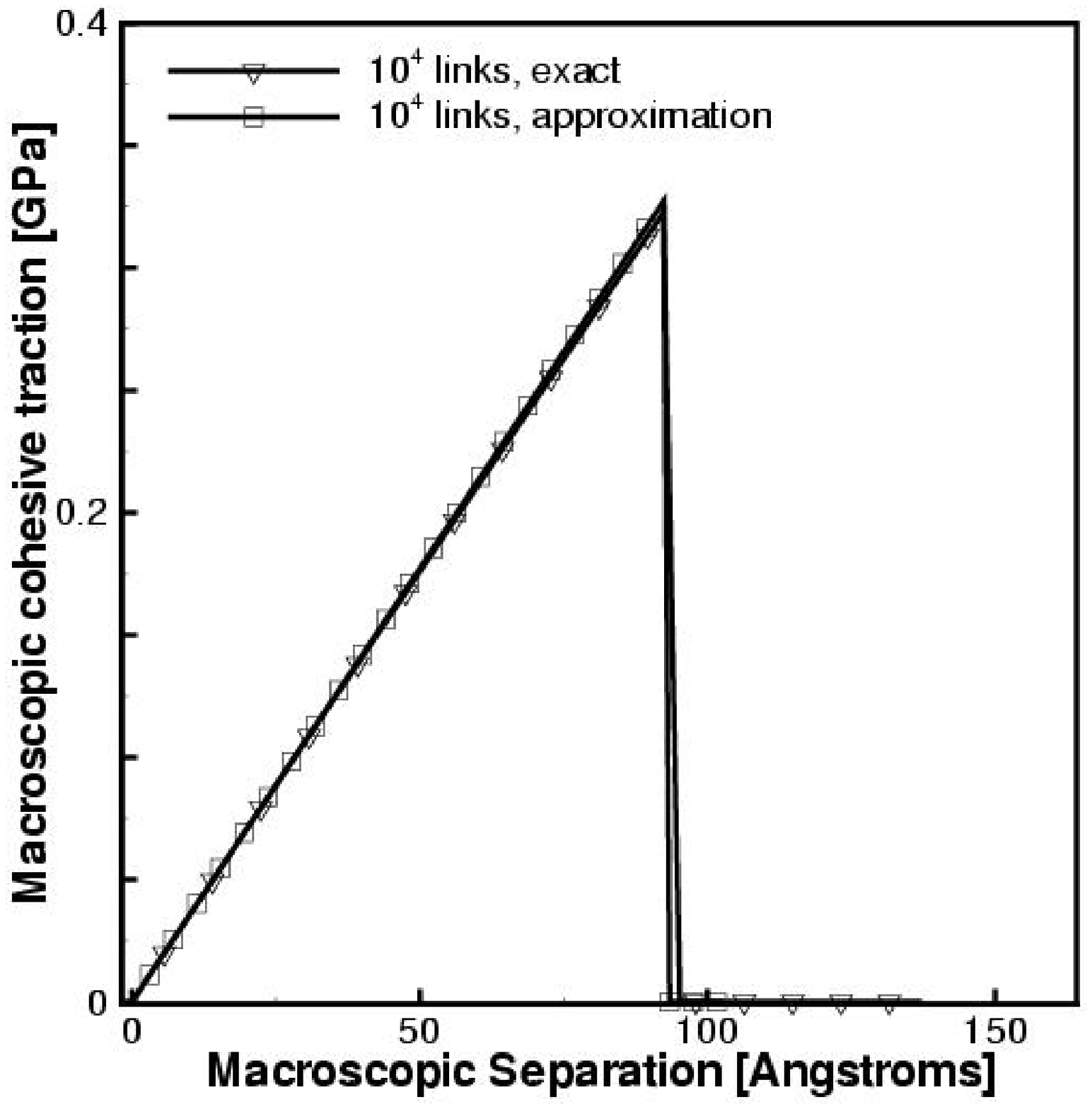}
    }
  \subfigure[$N=10^5$]{
    \label{ohtlinks}
    \includegraphics[width=.33\textwidth]{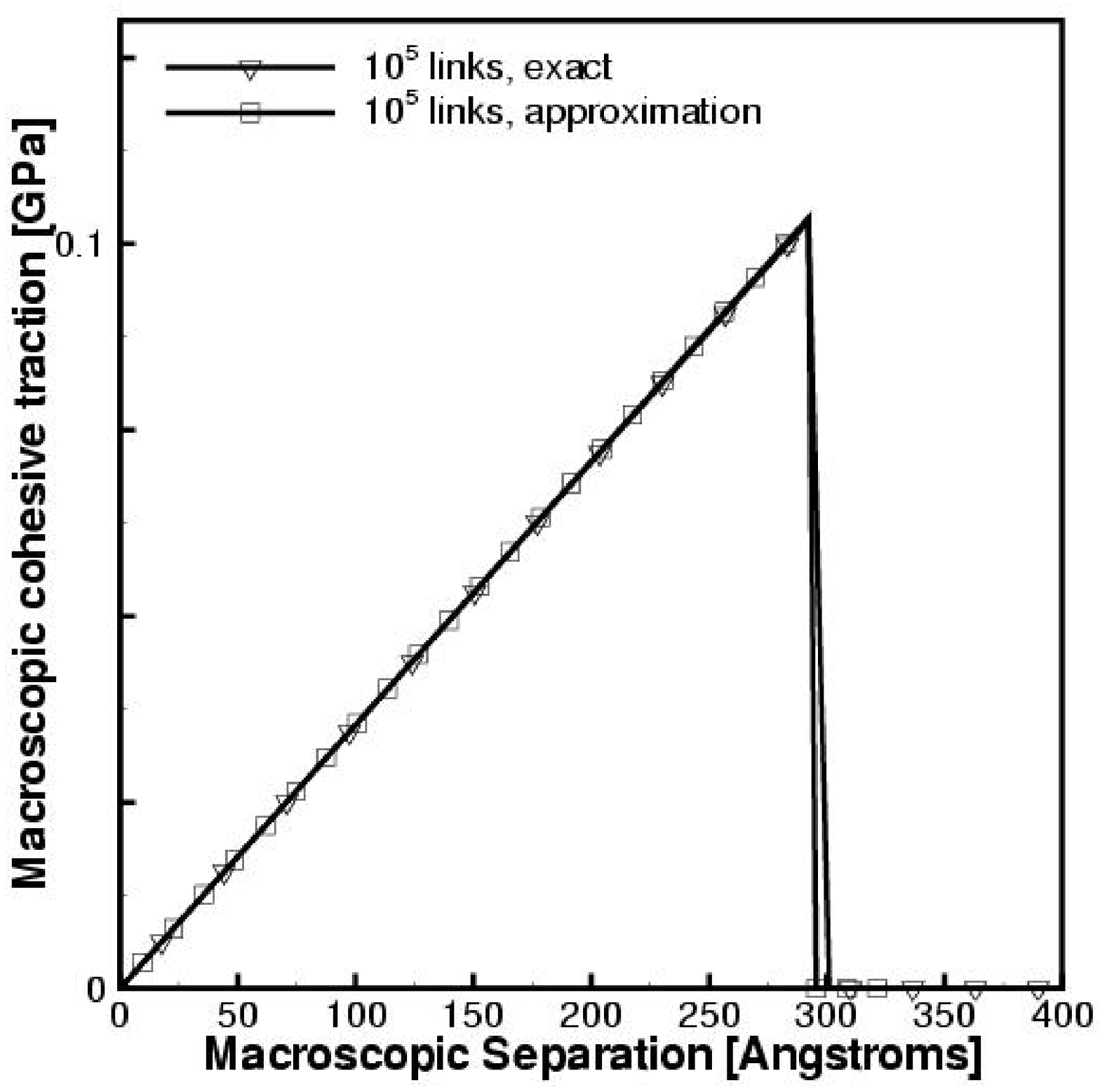}
    }
  \subfigure[$N=10^6$]{
    \label{omlinks}
    \includegraphics[width=.33\textwidth]{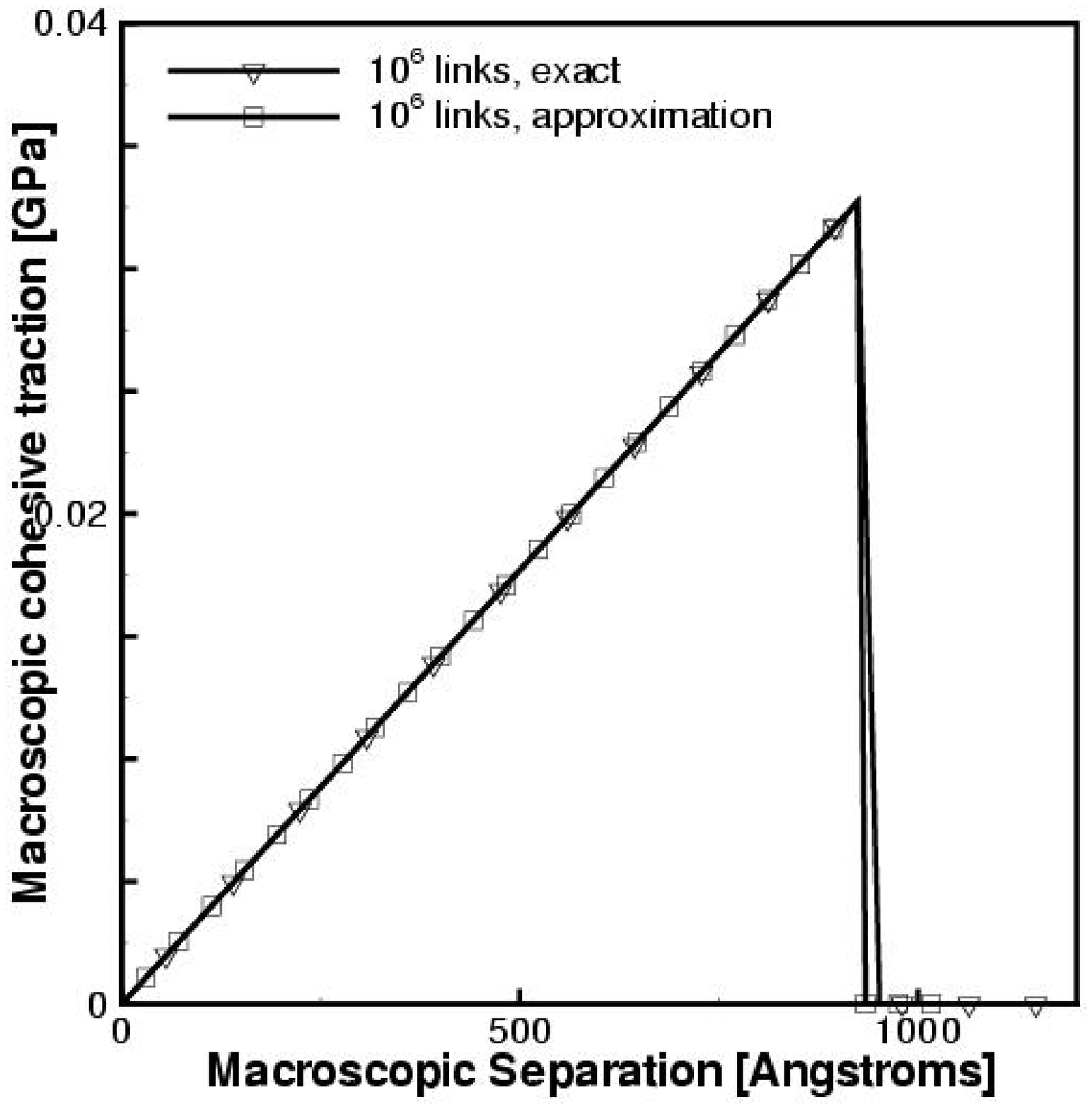}
    }
  \caption[]{Comparison between numerically computed macroscopic
  cohesive law and universal asymptotic form for an increasing number
  of atomic planes in the cohesive layer.}
\label{fig:comp_exact_asymp}
\end{figure}

As a simple illustrative example we consider the universal
binding energy relation (UBER) \cite{RoseSmithFerrante1983} defined by the
interplanar potential:
\begin{equation}\label{UBERpot}
  \phi(\delta)= 2\gamma - C \delta_c
  (\delta+\delta_c)e^{-\delta/\delta_c}
\end{equation}
This function falls within the class of potentials considered in
the foregoing. We choose as material constants: $C = 3.54$
J/m${}^2$/\AA${}^2$, $\delta_c = 0.66$ \AA, which are
representative of aluminum. The macroscopic cohesive laws
resulting from a direct numerical minimization of the energy
(\ref{Energy2}) for different values of $N$ are shown in
Fig.~\ref{fig:UBERtraction}. The universal asymptotic form of the
macroscopic cohesive law is compared in
Fig.~\ref{fig:comp_exact_asymp} against the corresponding
numerical results. The convergence of the macroscopic cohesive
law towards the universal asymptotic form as the number of planes
in the cohesive layer increases is clearly evident in this figure.

\section{Effect of finite temperature}

At finite temperature, entropic effects make it feasible for the
cohesive layer to decohere on multiple planes. In order to assess
this effect simply, we recall that, asympotically, the energy of
a cohesive layer with no decohered planes is $(\bar{C}/2)
\bar{\delta}^2$, and that the energy of a cohesive layer
containing $k$ decohered planes is $2k\gamma$. Within this
approximation, the partition function of an area $a^2$ of layer
is, therefore,
\begin{equation}\label{Z}
Z(\bar{\delta}, T) = {\rm e}^{-\beta (\bar{C}/2) \bar{\delta}^2
a^2} + \sum_{k=1}^\infty {\rm e}^{-\beta 2k \gamma a^2}
\end{equation}
where $\beta = 1/kT$, and $k$ is Boltzmann's constant. In order
to count states properly, we identify $a$ with the lattice
parameter of the crystal. Physically, this is tantamount to
allowing for decohered areas of a size commensurate with the
lattice parameter. The sum in (\ref{Z}) defines a geometric series
which may be evaluated readily, with the result:
\begin{equation}
Z(\bar{\delta}, T) = {\rm e}^{-\beta (\bar{C}/2) \bar{\delta}^2
a^2} + \frac{{\rm e}^{-\beta 2 \gamma a^2}}{1 - {\rm e}^{-\beta 2
\gamma a^2}}
\end{equation}
The free energy density per unit area of the layer now follows as
\begin{equation}
\bar{F}(\bar{\delta}, T) = - \frac{1}{a^2} \frac{1}{\beta} \log
Z(\bar{\delta}, T)
\end{equation}
whereas the resulting effective cohesive law is:
\begin{equation}
\bar{t}(\bar{\delta}, T) = \frac{d\bar{F}}{d\bar{\delta}}
(\bar{\delta}, T)
\end{equation}

\begin{figure}[h]\label{fig:exact_Statistical_mec}
  \centering
  \subfigure[Helmholtz free energy.]{
    \includegraphics[width=.64\textwidth]{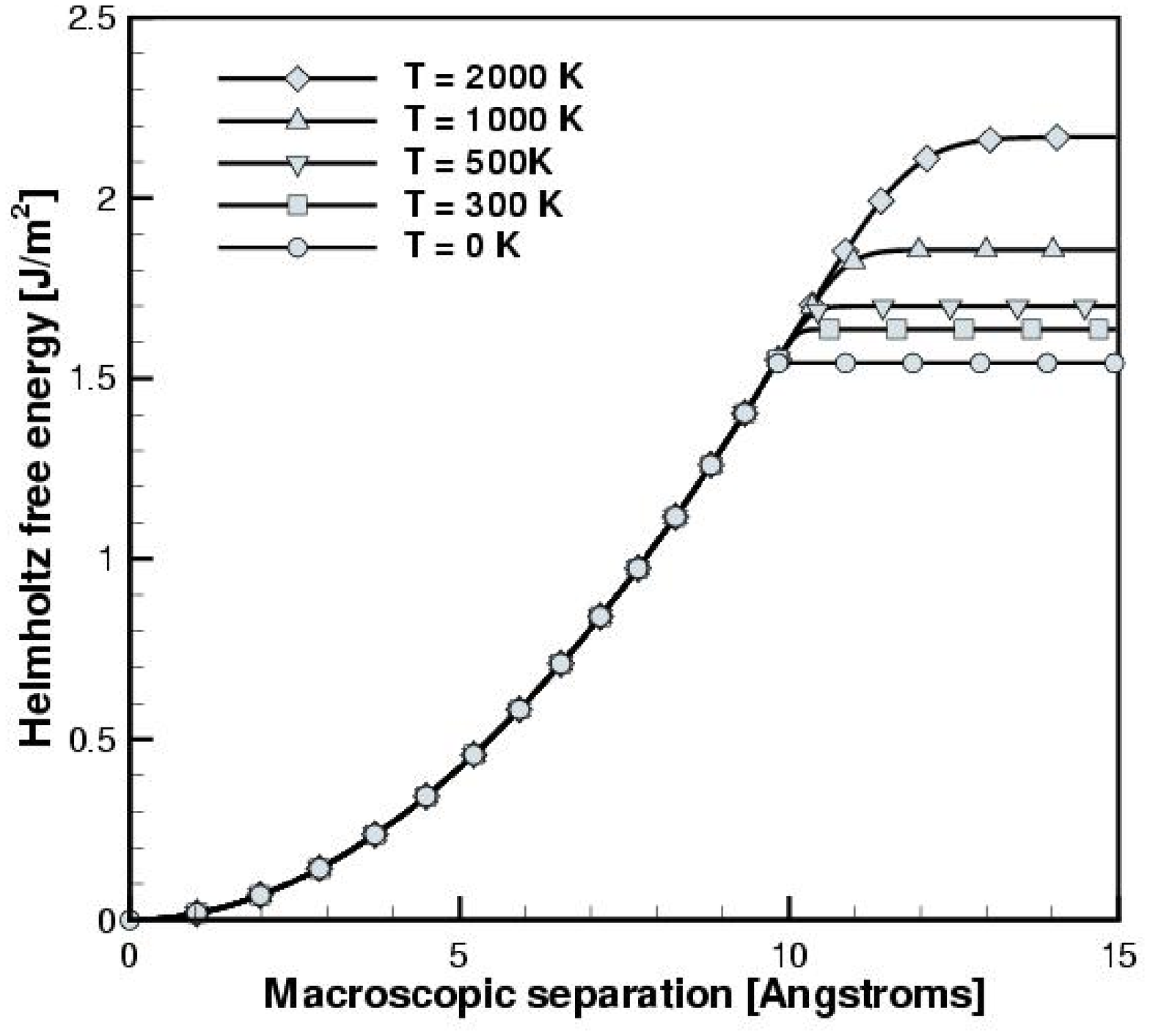}
    }
  \subfigure[Macroscopic cohesive traction.]{
    \includegraphics[width=.64\textwidth]{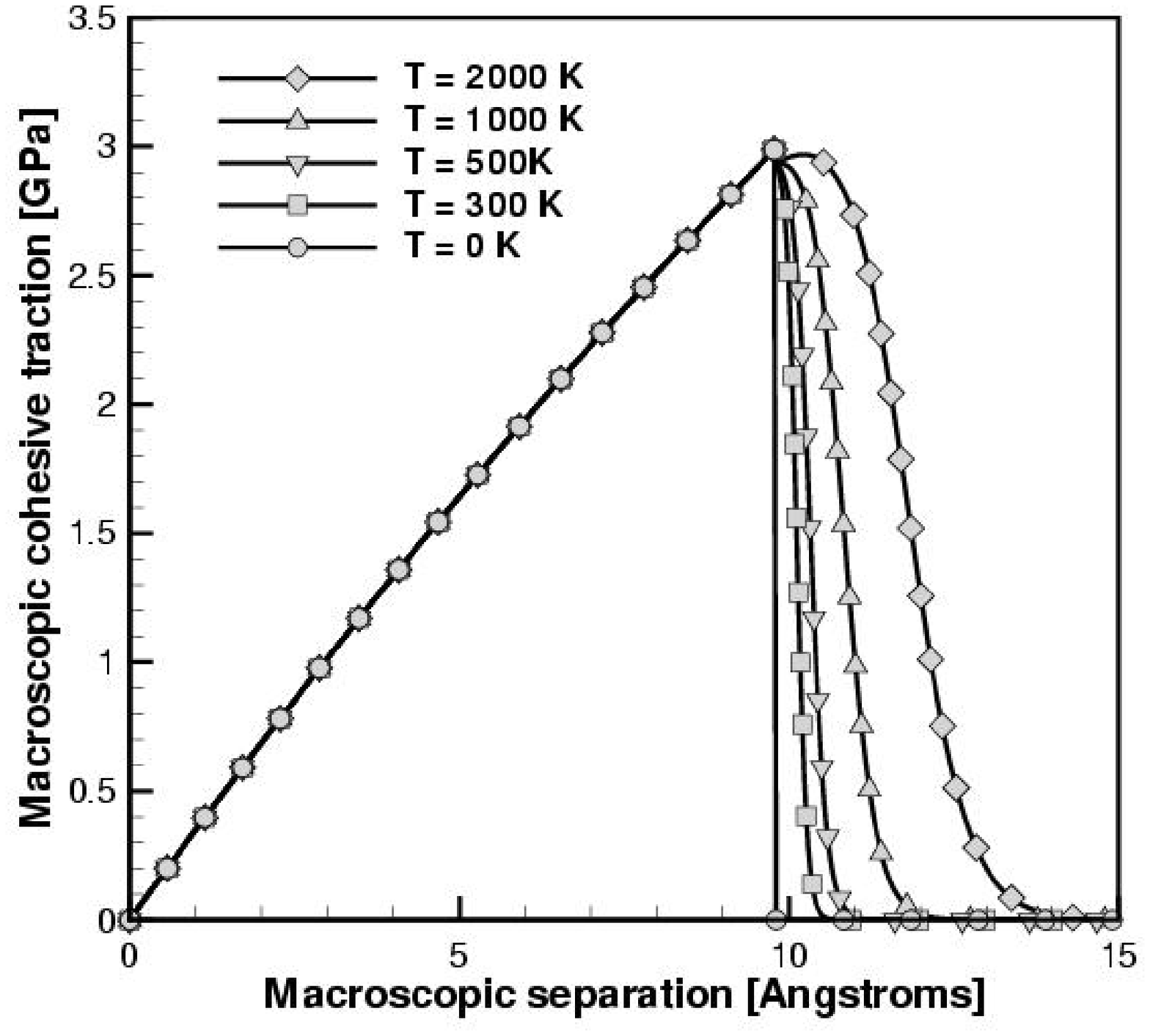}
    }
  \caption[]{Influence of the temperature on the effective behavior
    of the chain~($N=100$).}
    \label{fig:Temperature100}
\end{figure}

The effect of temperature on the macroscopic cohesive potential
for aluminum, endowed with an interplanar potential of the UBER
type, is shown in Fig.~\ref{fig:Temperature100}. The lattice
parameter is taken to be $a = 4.05$ \AA. As expected, the
Helmholtz free energy rises above the zero-temperature in the
amount $TS$, where $S$ is the configurational entropy of the
layer, Fig.~\ref{fig:Temperature100}a. The corresponding effect
on the macroscopic cohesive law is to smooth out the decohesion
transition, Fig.~\ref{fig:Temperature100}b.

\section{Renormalization Group interpretation}

The renormalization group (RG) (see, e.~g.,
\cite{Goldenfeld1992}) provides a natural framework for the
understanding of universality, i.~e., the phenomenon that large
classes of systems with unrelated Hamiltonians may nevertheless
exhibit identical thermodynamic behavior near critical points. It
is therefore not entirely unexpected that the main result of
Section~\ref{ZeroTemperature}, namely, that the limiting form of
the potential of a cohesive layer is universal for a broad class
of interplanar potentials, can be given a compelling
interpretation within the RG framework.

For simplicity, we confine our attention to the zero temperature
case. We proceed to construct an RG transformation $R$ such that
the sequence
\begin{equation}\label{Sequence}
\phi_{n+1} = R \phi_n, \quad n = 0, 1, \dots
\end{equation}
with $\phi_0(\delta) = \phi(\delta)$, yields, by recourse to an
appropriate scaling, the large $N$ asymptotic form of the
macroscopic cohesive law in the limit. As in the preceding
section, the interplanar potentials contemplated here, and to
which the transformation $R$ is applied, are continuous,
monotonically increasing functions $\phi:[0, \infty) \to [0, 2
\gamma]$ and analytic at the origin. For simplicity, we
additionally restrict attention to functions $\phi(\delta)$
possessing a single inflection point, so that $\phi'(\delta)$ has
a single maximum.

We construct $R$ by the usual combination of decimation and
scaling. The decimation step concerns a layer consisting of two
interatomic planes with opening displacements $\delta_1$ and
$\delta_2$ and total opening displacement $\delta$. The
corresponding effective energy follows from the minimization
problem:
\begin{eqnarray}
\tilde{\phi}(\delta) &=& \inf_{\{\delta_1, \delta_2\}} \{
\phi(\delta_1) + \phi(\delta_2) \} \label{RGEnergy} \\
\delta &=& \delta_1 + \delta_2 \label{RGKinematics}
\end{eqnarray}
Next, we proceed to rescale $\tilde{\phi}(\delta)$ in such a way
that the sequence defined by (\ref{Sequence}) has a well-defined
fixed point. Since the transformation must preserve the relation
$\phi(\infty) = 2\gamma$, it is clear that we are allowed to
rescale the independent variable $\delta$ only. Thus, we set:
\begin{equation}
(R\phi)(\delta) = \tilde{\phi}(\lambda \delta)
\end{equation}
For very small $\delta$ the interplanar potential $\phi$ is
essentially quadratic and reflects the elasticity of the lattice.
Thus, in order for the transformation to preserve the elasticity
of the lattice it must leave parabolic functions invariant. For
$\phi = (C/2)\delta^2$ it follows that
$\delta_1=\delta_2=\delta/2$ and $\tilde{\phi} =
2(C/2)(\delta/2)^2 = (C/2) (\delta/\sqrt{2})^2$. Finally,
$(R\phi)(\delta) = (C/2) (\lambda \delta/\sqrt{2})^2$, whence it
follows that $\lambda = \sqrt{2}$ for $\phi$ to remain invariant
under the transformation. The complete RG transformation is,
therefore,
\begin{eqnarray}
\tilde{\phi}(\delta) &=& \inf_{\{\delta_1, \delta_2\}} \{
\phi(\delta_1) + \phi(\delta_2) \} \label{RGEnergy2} \\
\delta &=& \delta_1 + \delta_2 \label{RGKinematics2} \\
(R\phi)(\delta) &=& \tilde{\phi}(\sqrt{2}\delta) \label{Scaling}
\end{eqnarray}
Taking $\delta_1=0$ and $\delta_2=\delta$ in (\ref{RGEnergy2})
immediately shows that the unscaled energy $\tilde{\phi}(\delta)$
is bounded above by the original function $\phi(\delta)$.

It is clear from definition (\ref{RGEnergy2}-\ref{Scaling}) that
the RG transformation leaves the specific fracture energy
$\phi(\infty)$ invariant and equal to its initial value $2\gamma$.
Another invariant of the RG transformation is the initial modulus
$C = \phi"(0)$. Indeed, consider the limit of $R\phi$ as
$\delta\to 0$. Since necessarily $\delta_1 < \delta$ and
$\delta_2 < \delta$ it follows that both $\delta_1\to 0$ and
$\delta_2\to 0$ in this limit. $(R\phi)(\delta)$ may therefore be
computed by replacing $\phi(\delta)$ by its Taylor expansion about
the origin, namely, $(C/2)\delta^2$, with $C=\phi"(0)$. But
parabolic functions are invariant under $R$ and, hence, so is $C$.

The RG transformation $R$ (\ref{RGEnergy2}-\ref{Scaling})
preserves the monotonicity of $\phi(\delta)$. In order to see
this, consider a pair of opening displacements $\delta$ and
$\delta' = \lambda \delta$, with $\lambda < 1$. Let
$\tilde{\phi}(\delta) = \phi(\delta_1) + \phi(\delta_2)$ for some
pair of opening displacements $\delta_1$ and $\delta_2$
satisfying constraint (\ref{RGKinematics2}). The opening
displacements $\delta'_1 = \lambda \delta_1$ and $\delta'_2 =
\lambda \delta_2$ then satisfy the similar constraint: $\delta' =
\delta'_1 + \delta'_2$. It therefore follows that
$\tilde{\phi}(\delta') < \phi(\lambda \delta_1) + \phi(\lambda
\delta_2) < \phi(\delta_1) + \phi(\delta_2) =
\tilde{\phi}(\delta)$. An application of the rescaling
(\ref{Scaling}) to both sides of this inequality finally proves
the assertion.

It is easy to show that the function:
\begin{equation}\label{FixedPoint}
\phi_\infty(\delta) = \min \{\frac{C}{2} \delta^2, 2\gamma \}
\end{equation}
is a fixed point of $R$. To this end, we may distinguish the
cases: a) $\delta_1<\delta_c$ and $\delta_2<\delta_c$; b)
$\delta_1<\delta_c$ and $\delta_2>\delta_c$, or, equivalently,
$\delta_2<\delta_c$ and $\delta_1>\delta_c$, and c)
$\delta_1>\delta_c$ and $\delta_2>\delta_c$. Case (a) requires
that $\delta < 2 \delta_c$ and gives an unscaled energy:
$\tilde{\phi}(\delta) = (C/2)(\delta/\sqrt{2})^2$. Case (b)
requires that $\delta > \delta_c$. The optimal unscaled energy is
obtained by taking $\delta_1=0$ and $\delta_2=\delta$, with the
result: $\tilde{\phi}(\delta) = 2\gamma$. Case (c) results in the
unscaled energy: $\tilde{\phi}(\delta) = 4\gamma$. The function
(\ref{FixedPoint}) is recovered  by taking the minimum of the
unscaled energies resulting from cases (a), (b) and (c) and
applying the scaling (\ref{Scaling}) to the result.

\begin{figure}[h]
  \begin{center}
    \epsfig{file=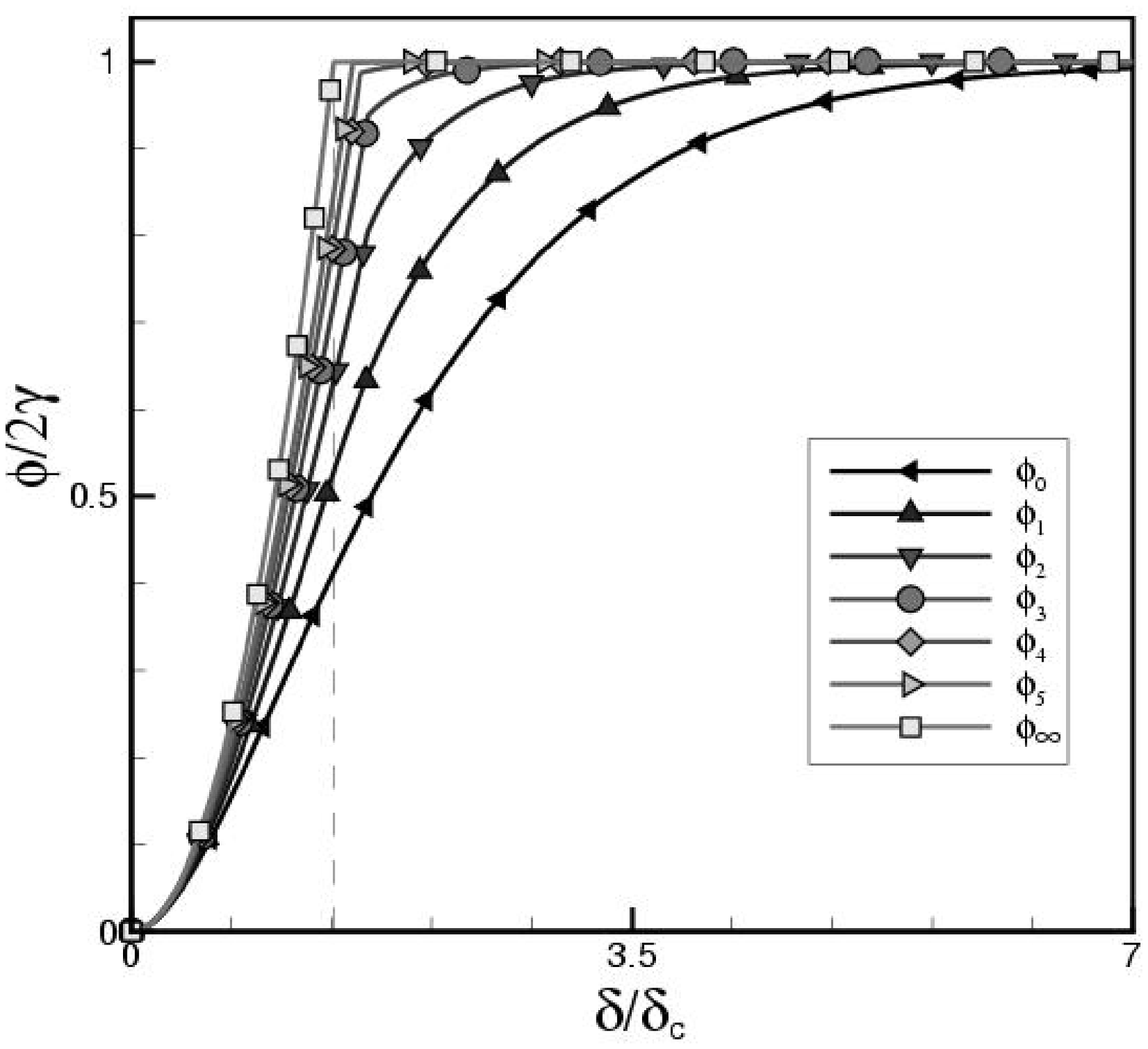,height=4.in}
    \caption[]{Evolution of the sequence $\phi_n,\ n = 0, 1, \dots$
      towards the {\it fixed point} $\phi_\infty$.}
      \label{fig:RGiterations}
  \end{center}
\end{figure}

A key question is whether the fixed point (\ref{FixedPoint}) is
attractive. We have investigated this question numerically for the
particular example of the UBER binding law, (\ref{UBERpot}).
Fig.~\ref{fig:RGiterations} shows the evolution of $\phi_n$ with increasing $n$.
It is clear from the figure that, at least for the example under
consideration, the flow of functions $\phi_n(\delta)$ does indeed
converge strongly to the fixed point (\ref{FixedPoint}).
The relation to the asymptotic cohesive law (\ref{AsymptoticPhi})
is as follows. We may regard $\phi_n(\delta)$ as the result of
decimating and rescaling $n$ times a cohesive layer containing $N
= 2^n$ planes. The total opening displacement of the layer is
obtained by undoing all the rescalings, with the result:
$\bar{\delta} = (\sqrt{2})^n \delta = \sqrt{N}\delta$. For large
$N$, $\phi_n(\delta) \sim \phi_\infty(\delta)$ and one has
\begin{equation}\label{RGvsAsymptotic}
\bar{\phi}(\bar{\delta}) \sim \phi_\infty(\bar{\delta}/\sqrt{N}),
\qquad \text{as } N \to \infty
\end{equation}
which is identical to (\ref{AsymptoticPhi}).

Eq.~(\ref{RGvsAsymptotic}) establishes a connection between the
renormalization group, specifically as generated by transformation
(\ref{RGEnergy2}-\ref{Scaling}), and the large-$N$ asymptotic form
of the cohesive potential determined directly in the preceding
section. It is interesting to note that the RG transformations
which pertain to the renormalization of interplanar potentials
are markedly different from those which arise in the calculation
of bulk thermodynamic properties. In this latter context, the
appropriate scaling is related to the volume of the sample and is
designed so as to result in well-defined extensive fields and
intensive variables. In the present context, the energy densities
under consideration are defined per unit area, and the limit of
interest is the total energy of the cohesive layer per unit
surface area, as opposed to the energy per unit volume. In
addition, the independent variable of interest is the total
opening displacement across the cohesive layer, as opposed to its
transverse strain. These peculiarities account for the
non-standard character of the renormalization group defined in
the foregoing.

\section{Summary and conclusions}


We have presented two approaches for coarse-graining interplanar
potentials and determining the corresponding macroscopic cohesive
laws based on energy relaxation and the renormalization group. We
have analyzed the cohesive behavior of a large---but
finite---number $N$ of interatomic planes and found that the
macroscopic cohesive law adopts a universal asymptotic form in
the limit of large $N$. We have also found that this asymptotic
form of the macroscopic cohesive law is an attractive fixed point
of a suitably-defined renormalization-group transformation.

The universal asymptotic form of cohesive law is particularly
simple: the traction rises linearly from zero to a peak stress
$\bar{\sigma}_c$ at a critical opening displacement
$\bar{\delta}_c$, and subsequently drops to zero. The scaling of
the peak stress and critical opening displacement is
$\bar{\delta}_c \sim 1/\sqrt{N}$ and $\bar{\delta}_c \sim
\sqrt{N}$. Thus, coarse-graining is accompanied by an attendant
reduction (increase) in the cohesive traction (opening
displacement) range, while at the same time preserving the
surface of specific fracture energy of the crystal.

It is interesting to note that the size of a cohesive zone at the
tip of a crack in an elastic crystal scales as $l \sim
1/\sigma_c^2$ \cite{barrenblatt:1962}. Upon coarse-graining, the cohesive zone
size increases to $\bar{l} \sim 1/\bar{\sigma_c}^2 $, which gives
$\bar{l} \sim N l$. This scaling preserves the ratio $\bar{l}/l =
\bar{d}/d$, which shows that coarse-graining has the effect of
expanding the cohesive-zone size to within the resolution of the
coarse-grained description. In particular, it eliminates the
onerous need to resolve the atomic scale in simulations.

It is also interesting to note that the universal form of the
macroscopic cohesive potential is completely determined by the
constants $C = \phi''(0)$ and $\phi(\infty) = 2\gamma$. This
greatly reduces the scope of the first-principles calculations
required to identify the macroscopic cohesive behavior of specific
materials, which can be limited to the calculation of elastic
moduli, lattice constants and surface energies.

Finally, we close by suggesting possible extensions of the theory.
The analysis presented in the foregoing has been restricted to
symmetric (mode I) opening normal to the atomic planes. A
worthwhile extension would be to consider interplanar potentials
defined in terms of three opening displacements and, therefore,
capable of describing tension-shear coupling. Another worthwhile
extension would be to consider interplanar potentials with
multiple inflection points, which would greatly enlarge the class
of materials tractable within the theory.

\section*{Acknowledgements}

This work has been supported by Brown University's MURI Center
for the ``Design and Testing of Materials by Computation: A
Multi-Scale Approach." We are grateful to Emily A.~Carter, Emily
A.~A.~Jarvis and R.~L.~Hayes for many useful discussions and
suggestions, and for making available to us their research
results prior to publication.

\bibliographystyle{plain}
{\small\bibliography{RNG}}

\end{document}